\documentclass[12pt,preprint]{aastex}
\usepackage{epstopdf}
\newcommand{\teff}{$T_{\textrm{eff}}$}

\shorttitle{Tracing the origin of the $\gamma$ Leo stream}
\shortauthors{Liang et al.}

\begin{document}
\title{Tracing the origin of moving groups\\
\uppercase\expandafter{\romannumeral1}. The $\gamma$ Leo moving group with high resolution spectra from the Subaru Telescope }
\author{X. L. Liang\altaffilmark{1,2}, J. K. Zhao\altaffilmark{1}, G. Zhao\altaffilmark{1,2}, W. Aoki\altaffilmark{3}, M. N. Ishigaki\altaffilmark{4}, T. Matsuno\altaffilmark{3}, Y. Q. Chen\altaffilmark{1,2}, X. M. Kong\altaffilmark{1}, J. R. Shi\altaffilmark{1,2}, Q. F. Xing\altaffilmark{1}}

\altaffiltext{1}{Key Laboratory of Optical Astronomy, National Astronomical Observatories, Chinese Academy of Sciences, Beijing 100012, China. zjk@nao.cas.cn}
\altaffiltext{2}{School of Astronomy and Space Science, University of Chinese Academy of Sciences, Beijing 100049, China}
\altaffiltext{3}{National Astronomical Observatory of Japan, 2-21-1 Osawa, Mitaka, Tokyo 181-8588, Japan; Department of Astronomical Science, School of Physical Sciences, The Graduate University of Advanced Studies (SOKENDAI), 2-21-1 Osawa, Mitaka, Tokyo 181-8588, Japan. aoki.wako@nao.ac.jp}
\altaffiltext{4}{Kavli Institute for the Physics and Mathematics of the Universe (WPI), Todai Institute for Advanced Study, University of Tokyo, 5-1-5 Kashiwanoha, Kashiwa, 277-8583 Chiba, Japan. miho.ishigaki@ipmu.jp}

\begin{abstract}
We present chemical abundances of 15 stars in the $\gamma$ Leo moving group based on high-resolution spectra with the Subaru High Dispersion Spectrograph. The sample was picked up by applying wavelet transform to UVW velocity components of stars in the solar neighbourhood. Both photometric and spectroscopic method have been used to determine the stellar parameters of stars. Abundances of 11 elements including Na, Mg, Al, Si, Ca, Ti, Cr, Fe, Ni, Y and Ba are measured. Our results show that the member stars display a wide metallicity distribution with abundance ratios similar to Milky way disk stars. We presume that the $\gamma$ Leo moving group is originated from dynamical effects probably related to the Galactic spiral arms.

\end{abstract}


\section{INTRODUCTION}

More and more complex substructures have been discovered in the Milky Way by recent digital sky surveys(\citet{ant12}; \citet{kle08}; \citet{zhao09}). It is well-known that the vicinal velocity field is clumpy and most of the observed overdensities are made of spatially unbound groups of stars, called moving groups. \citet{egg58} have defined and investigated many moving groups, supposing that moving groups are from dissolving open clusters. Later, many theoretical models suggest that the overdensities of stars in some regions of the Galactic velocity UV-plane may be a result of global dynamical mechanisms related to the nonaxisymmetry of the Galaxy \citep{fam05}, namely the presence of the bar \citep{kal91,deh00,fux01}, and/or spiral arms \citep{sim04,qui05}. Since the late 90s of last century, the bar has been believed to be short and fast rotating for long time. This was in very good agreement with the explanation of the Hercules moving group as being due to the bar's outer Lindblad resonance \citep{deh00}. However, recent photometric studies of the Galactic center have shown that the bar could be longer, reopening the debate on the bar's pattern speed and the origin of the Hercules moving group \citep{mon17,per17}. Nowadays, the origins of these moving groups are explained by different theories or hypotheses, such as cluster disruption, dynamical effects and accretion events. \citet{fre02} put forward the chemical tagging technique to reassemble the ancient stellar forming aggregates in the Galactic disk. Since then it has become popular to use detailed chemical abundances from high resolution spectroscopy to disentangle the mechanism that has formed a certain stream. For example, \citet{ben07} found a wide spread in the distributions of age and chemical abundances of the stars in the Hercules stream, and concluded that this group is compatible with being a dynamical feature. According to the homogeneity of the HR 1614 group in age and abundance, \citet{sil07} concluded that it is the remnant of a dispersed star-forming event.

In the past, it is hard to determine the stellar members of moving groups due to the lack of parallaxes information, which will become available with Gaia survey. Combined with spectroscopic survey, like LAMOST, we can determine accurate velocity coordinates of stars in the solar neighbourhood. The $\gamma$ Leo (Leonis) moving group hasn't been closely analysed before. The $\gamma$ Leonis group were defined by \citet{egg59a,egg59b} by convergent point method. Its existence has been confirmed by \citet{sku97}. \citet{ant12}, using RAVE data, reidentified two peaks of the $\gamma$ Leo moving group in UV plane by wavelet transform which are confirmed by \citet{lia17} using Gaia-TGAS (\citet{pru16,bro16}) cross-matched with LAMOST DR3 \citep{cui12,zhao06,zhao12}.

The objective of this paper is to trace the origin of the $\gamma$ Leo moving group by chemical tagging. Section 2 describes our sample and observational information about this sample. In Section 3, we discuss stellar parameters, chemical abundance and error analysis. The main results and discussions are given in Section 4. In the final section, we present conclusion of our work and expectation for the future.

\section{SAMPLE SELECTION and OBSERVATION}

\subsection*{Membership Criteria}
Membership of a moving group is based on the stars' velocities. The velocity components \textit{UVW} are defined in a right-handed local standard reference coordinated system, which point to the directions of the Galactic centre, Galactic rotation and the North Galactic Pole, respectively. Velocities were corrected to the local standard of rest where the Sun's \textit{UVW} velocities are ($7.01$, 10.13, 4.95) km s$^{-1}$ \citep{hua15} respectively. Parallaxes and proper motions were taken from Gaia DR1 (\citet{bro16, pru16}) and radial velocities were taken from the LAMOST catalogue (Zhao et al. 2012; Cui et al. 2012). With the correlation coefficients provided by Gaia, the uncertainties of the velocity components have been calculated using the full covariance matrix. We have excluded stars with relative parallax uncertainty larger than 30\% from the sample.

\citet{lia17} used a wavelet transform technique to search for overdensities in the velocity distribution. Following \citet{ant12}, peak 7 and peak 10 were identified as $\gamma$ Leo moving group in the local sample of \citet{lia17}. Wavelet transform was applied to the \textit{UVW} coordinates to get the peaks and the size of the $\gamma$ Leo overdensity. Then we took the peak (60.8, 3.3, 2.9) km s$^{-1}$ as the center and the spherical radius 9.3 km s$^{-1}$ as the radius of the $\gamma$ Leo moving group. We adopted all objects within the radius as candidate stars, taking account of typical errors of velocity about 5 km s$^{-1}$ and inner velocity dispersion of a moving group is more than 2 km s$^{-1}$ \citep{shk12}. 77 candidate stars within 300 pc were selected, from which 18 stars have been observed with the Subaru Telescope. Table 1 lists their identifier; equatorial coordinate; parallax; proper motion components and identifier name in the simbad astronomical database of 15 single stars (other 3 stars are spectroscopic binaries).

\subsection*{Observation}
High-resolution spectra of 18 candidates of the $\gamma$ Leo moving group members were obtained on August 3-6, 2017 with Subaru/HDS (High Dispersion Spectrograph; \citealt*{nog02}). The spectral resolution is 36,000 covering approximately from 4000 {\AA} to 7000 {\AA}. The data were reduced by IRAF echelle package, including bias correction, flat fielding, scattered light subtraction, extraction of spectra, and wavelength calibration using Th arc lines. Cosmic-ray hits are removed by the method described in \citet{aok05}. The code HDSV \citep{zhao07} were used to estimate heliocentric radial velocities and normalize the spectra. The signal to noise ratio (\textit{S/N}) of spectra varies from star to star (listed in table~\ref{tbl-2}), but the mean value is 68.2 per pixel at 5500 {\AA}. Among the 18 observed stars, three were found to be double lined spectroscopic binaries and they were excluded from the sample. The remaining 15 stars have been analyzed. Solar spectrum (moon spectrum) observed by NAOC-Xinglong 2.16 m Telescope was acquired for correcting the zero point of elemental abundances.

\section{ABUNDANCE ANALYSIS}

\subsection*{Equivalent width measurements and Model Atmospheres} \label{bozomath}
The elemental abundances were determined based on equivalent widthes (EWs) measured line by line with gaussian fit. The atomic data of all the lines we used are taken from \citet{kon17}. To estimate the accuracy of EWs measurement, we compared our EWs of the moon spectrum with those of \citet{ben14}. The linear least squares fit for the two sets of data is $EW_{this work}=1.0225(\pm0.0054)EW_{Bensby}+1.055(\pm0.351)$ m\AA, and the standard deviation is about 2.3 m\AA. Figure 1 shows a comparison of the two EWs sets and their linear fit line. The model atmospheres were interpolated from LTE Kurucz model atmospheres \citep{kur93} and the theoretical EWs for individual lines were calculated using the ABONTEST8 code supplied by Dr. P. Magain.
\begin{figure}
\epsscale{.90}
\plotone{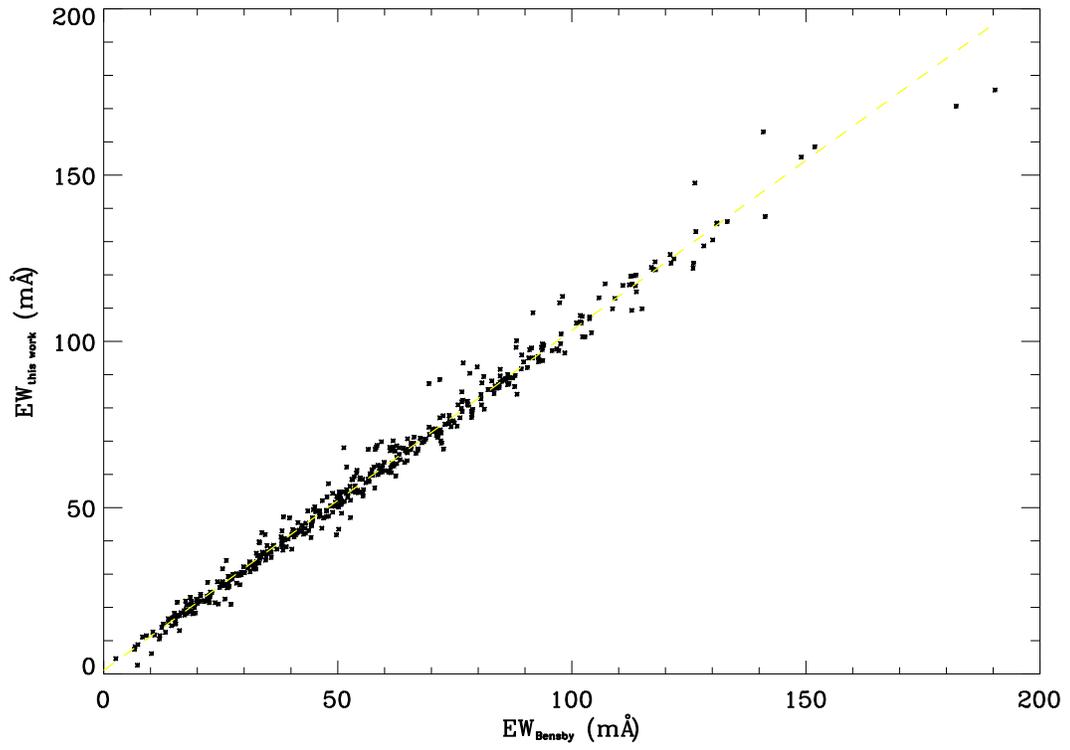}
\caption{Equivalent widths comparison for sun between the values of \citet{ben14} and those measured in this work (x and y axes, respectively). The dashed line is a linear fit to the data points.\label{Figure 1}}
\end{figure}

\subsection*{Stellar Parameters} \label{bozomath}
Stellar parameters are estimated by two methods, photometric way and spectroscopic way. For the photometric way, effective temperatures (\teff) were derived from the photometric colour index \textit{V $-$ K} according to empirical calibration relations given by \citet{alo96}. The apparent magnitudes are adopted from SIMBAD Astronomical Database \citep{sbd00}. Most stars' color excess E(B $-$ V) are obtained from Galactic Dust Reddening and Extinction website\footnote{http://irsa.ipac.caltech.edu/applications/DUST/}\citep{sch11}. However, for J0219+5623, suspect that the value 0.43 obtained by this method is an overestimate and take 0.03 from 3D Dust Mapping website\footnote{http://argonaut.skymaps.info/}\citep{gre15} as E(B $-$ V) to estimate this star's effective temperature. Surface gravity ($\log g$) is calculated from basic relation between bolometric magnitude, effective temperature, stellar mass and surface gravity.
$$
\log\frac{g}{g_{\bigodot}} = \log\frac{M}{M_{\bigodot}}+4\log\frac{T_{eff}}{T_{eff,\bigodot}}+0.4(M_{bol}-M_{bol,\bigodot})
$$
where
$$
M_{bol} = V_{mag}+BC+5\log\pi+5
$$
The parallax $\pi$ (mas) is taken from Gaia-Tgas \citep{bro16, pru16}. The bolometric corrections are calculated using the relation given by \citet{alo95}. Most of our stars are turn-off stars, while for two sub-giant stars (J1735+2650 and J0159+2636), we used formulae provided by \citet{alo99} to calculate the effective temperature. Stellar mass is estimated by interpolation of the evolutionary tracks of \citet{yi03} for \teff and $M_{bol}$.

For the spectroscopic way, effective temperature is determined by adjusting excitation equilibrium, requiring the slopes of lower excitation potential vs log $\epsilon$(Fe \uppercase\expandafter{\romannumeral1}) close to zero. Surface gravity ($\log g$) is determined from ionization equilibrium method which forces $\log \epsilon$(Fe \uppercase\expandafter{\romannumeral1}) equal to $\log \epsilon$(Fe \uppercase\expandafter{\romannumeral2}). Micro-turbulence is determined as the abundances of Fe \uppercase\expandafter{\romannumeral1} lines show no trend with EWs. We iterate the fitting with a 3$\sigma$ rejection of the deviant Fe lines after the first determination of the stellar parameters. The parameters from this spectroscopic method are adopted as our final parameters to calculate the abundances. In figure 2, we plot stars' positions in HR diagram. The abscissa labels spectroscopic effective temperature and the ordinate labels luminosity. The dotted line is a Y$^2$ isochrone with age = 1.2 Gyr from \citet{yi01}.

\begin{figure}
\epsscale{.60}
\plotone{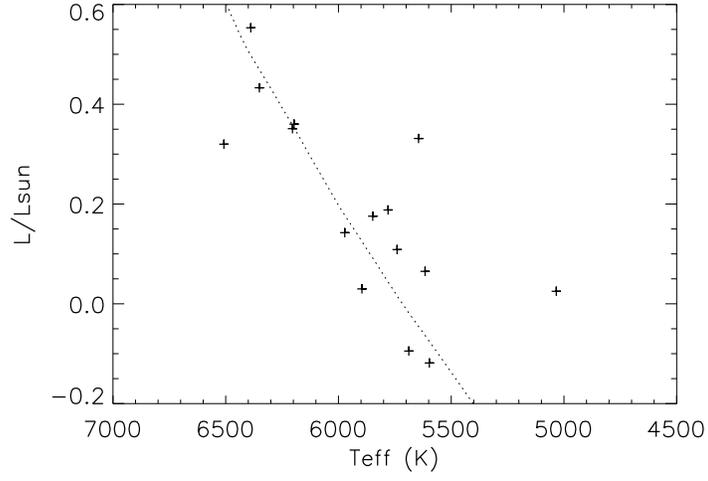}
\caption{HR diagram of our sample.\label{Figure 2}}
\end{figure}

The resulting stellar parameters are listed in table~\ref{tbl-2}. Figure 3 shows comparisons of \teff and $\log g$ from the two methods. The average and standard deviation of $\triangle$\teff are 66 K and 270 K, respectively. The systematic effect between these methods may be mainly caused by uncertainties of extinction. The average and standard deviation of $\triangle\log g$ are 0.007 and 0.069, respectively. We suppose the systematic deviation for $\log g$ can be ignored. The uncertainties of $\log g$ are mainly caused by uncertainties of stellar effective temperature and uncertainties of the mass from parallax errors and extinction.

\begin{figure}[htbp]
\begin{center}
\begin{minipage}[c]{0.4\textwidth}
\centering
\includegraphics[scale=0.35]{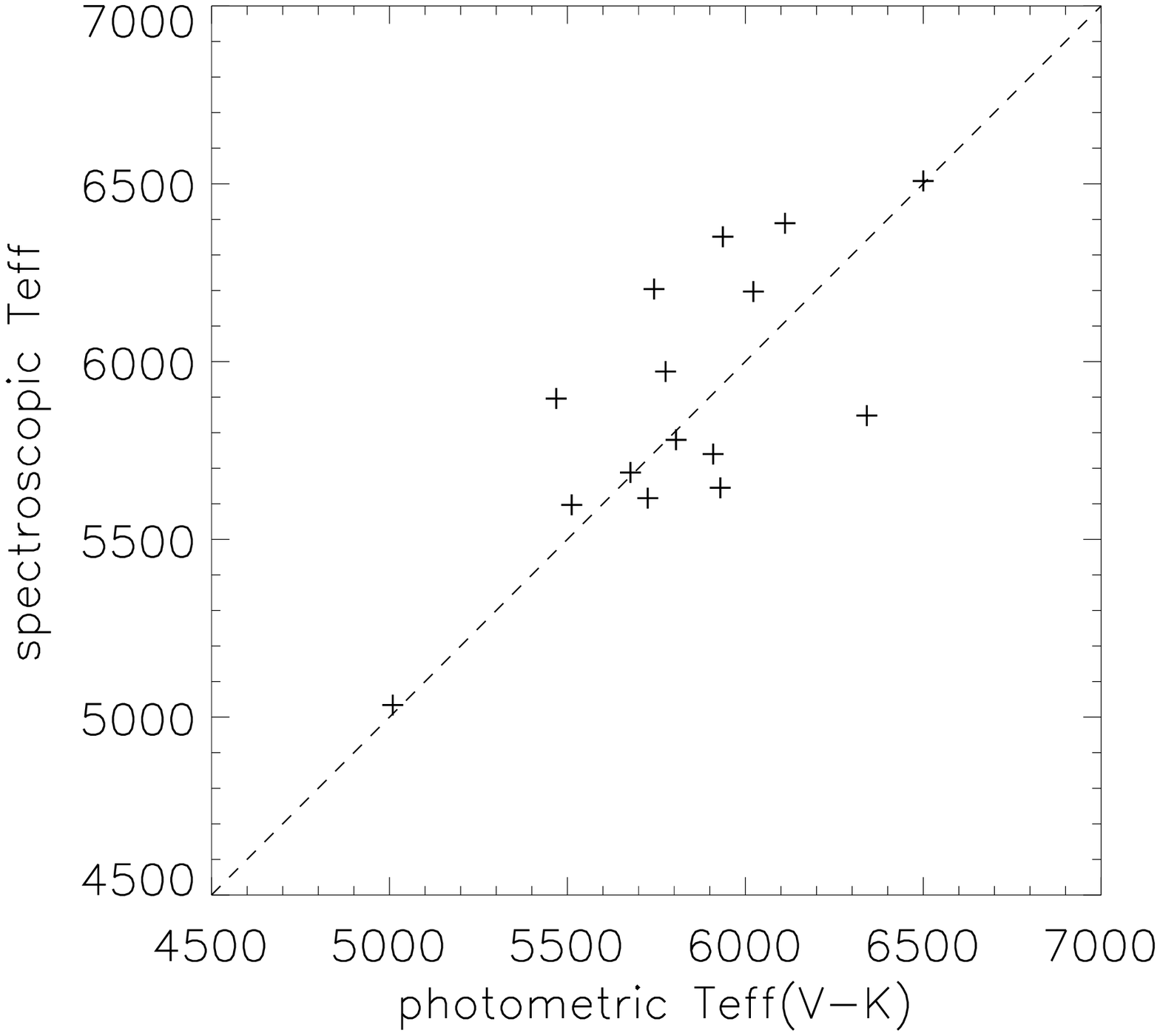}
\end{minipage}
\begin{minipage}[c]{0.4\textwidth}
\centering
\includegraphics[scale=0.35]{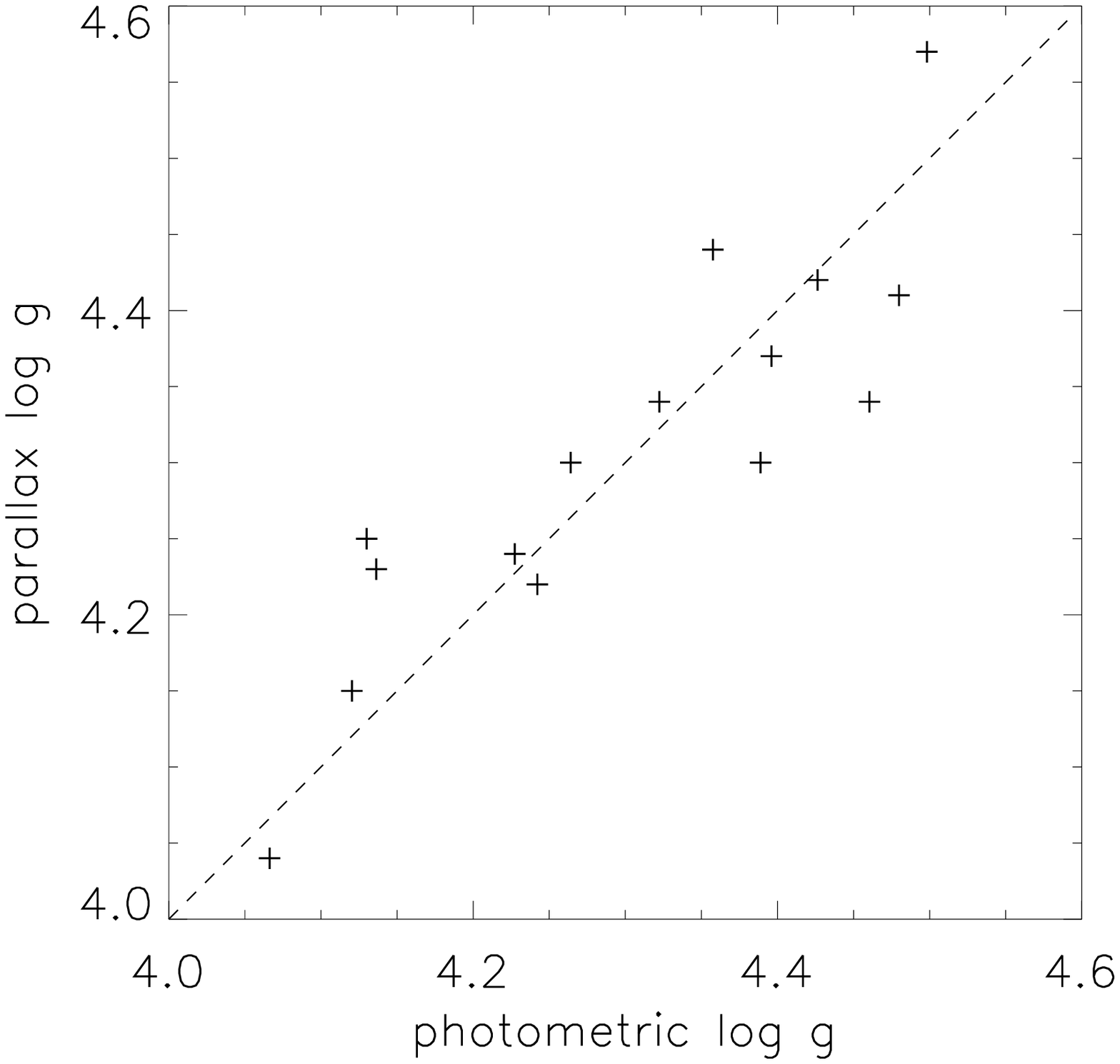}
\end{minipage}
\caption{Photometric \teff vs spectroscopic \teff and photometric $\log g$ vs spectroscopic $\log g$. The dash line indicates unit slope. \label{Figure 3}}
\end{center}
\end{figure}

We adopted the photospheric solar abundance of \citet{asp09}, to calculate [X/H] values. To estimate the offset of the derived abundances with respect to their results, we used the moon spectrum to gain the solar parameter. The gained effective temperature, surface gravity, micro-turbulent velocity and metallicity are \teff$ = $5779 K, $\log g = $4.35, $\xi = $0.85 km s$^{-1}$ and $\log \epsilon(Fe) = $7.61, respectively. Thus when calculated the metallicity [Fe/H] of other stars, we subtracted a extra 0.11 dex for system correction. The final stellar abundances in the [X/H] are presented in table~\ref{tbl-3} and abundance distributions are plotted in Figure 4, -5. Moreover, we overploted the abundances of field stars from \citet{ben14} and \citet{ven04} as comparison stars.

\subsection*{Error Analysis} \label{bozomath}

To estimate abundance uncertainties due to errors associated with EWs measurements and stellar parameters, we analyzed the sensitivities of abundance to changes of each quantity separately with the others unchanged. Table~\ref{tbl-4} lists the abundance differences induced by changing the equivalent widths $\Delta EW = 2.3$ m\AA, effective temperature $\Delta T_{\textrm{eff}}=100 $K, the surface gravity $\Delta\log g = 0.12$, the iron abundance $\Delta$[Fe/H]$=0.11$ dex and the micro-turbulent velocity $\Delta \xi = $0.1 km  s$^{-1}$, respectively. We took $\Delta EW = 2.3$ m\AA because the standard deviation of our measured EWs of the moon spectrum with those of \citet{ben14} is about 2.3 m\AA. The typical error in the stellar parameters $\Delta T_{\textrm{eff}}$, $\Delta\log g$ and $\Delta \xi$ are estimated based on our spectroscopic derivation, namely according to the slope changes. We took the $\Delta$[Fe/H]$=0.11$ dex because the maximum random error of [Fe/H] in the measurements is about 0.11 dex. Finally, we adopt the square root of the quadratic sum of the errors of all factors as the total error $\sigma_{total}$. We did not consider the NLTE effects which may cause larger scatter and overestimated abundances. As Table~\ref{tbl-4} shows, apart from the total error of Na abundance of a star (J0158+3955) that reaches 0.19 dex, the largest uncertainty is 0.14 dex. The titanium is more sensitive to changes of parameters than other elements, and the largest abundance error appears in the $\Delta$[Ti] column for most stars. These uncertainties do not change the result of abundance distribution.

\section{RESULT and DISCUSSION}

Figure 4 shows the abundance ratios of these elements for our sample and comparison stars. The metallicity of the 15 $\gamma$ Leo moving group member stars ranges from -0.67 to 0.35. The mean value and standard deviation are respectively 0.03 and 0.24. The large dispersion demonstrates they are not from a chemically homogeneous origin.

\begin{figure}[htbp]
\begin{center}
\begin{minipage}[c]{0.45\textwidth}
\centering
\includegraphics[scale=0.6]{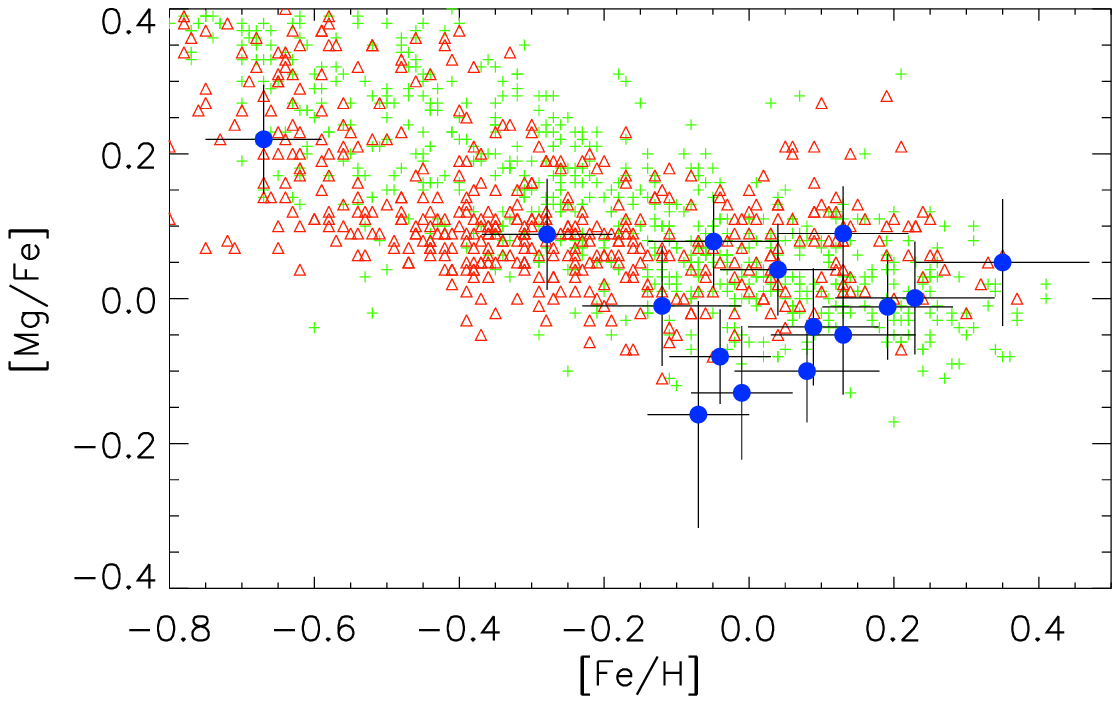}
\end{minipage}
\begin{minipage}[c]{0.45\textwidth}
\centering
\includegraphics[scale=0.6]{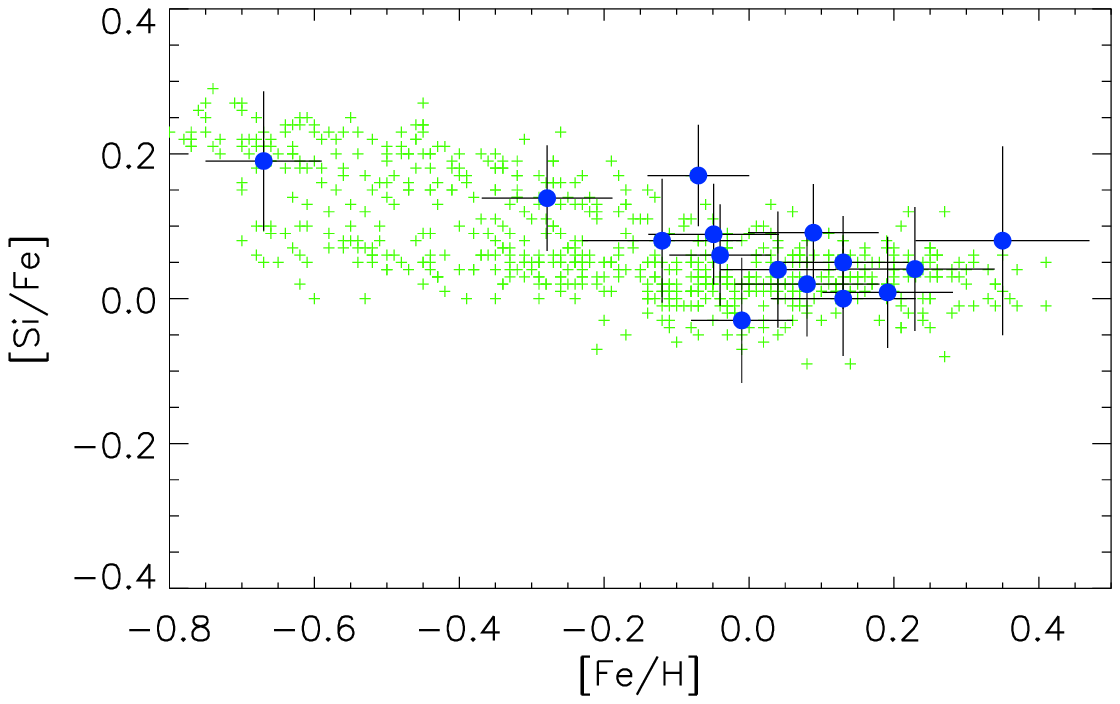}
\end{minipage}
\end{center}
\vspace{0pt}
\begin{center}
\begin{minipage}[c]{0.45\textwidth}
\centering
\includegraphics[scale=0.6]{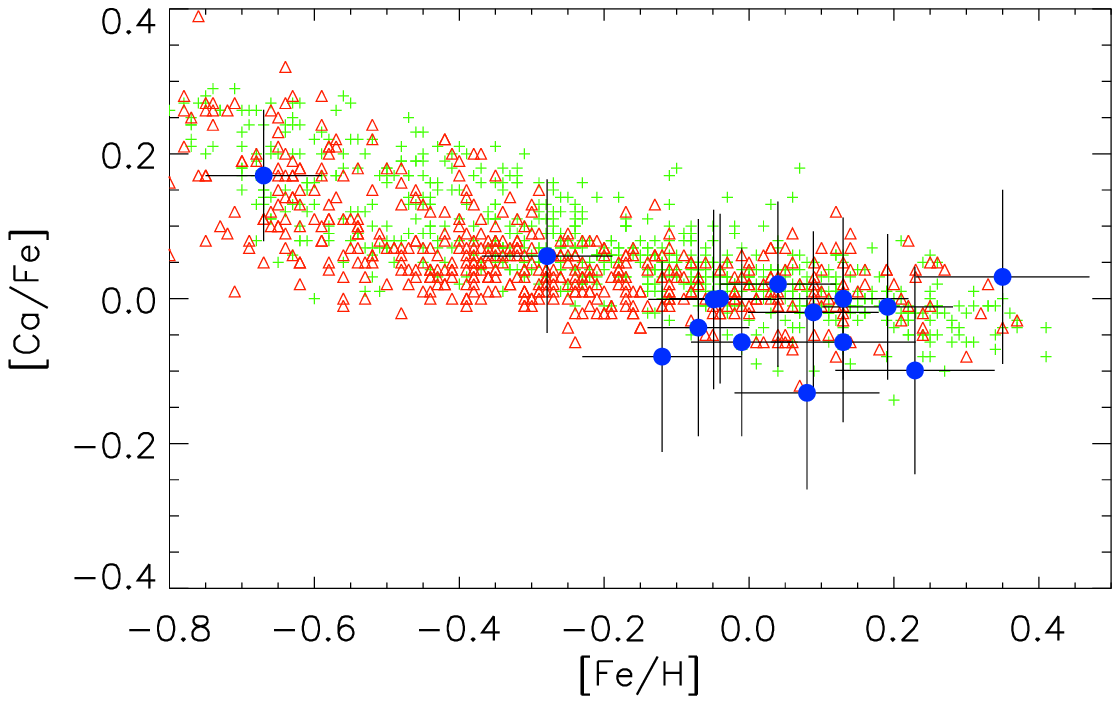}
\end{minipage}
\begin{minipage}[c]{0.45\textwidth}
\centering
\includegraphics[scale=0.6]{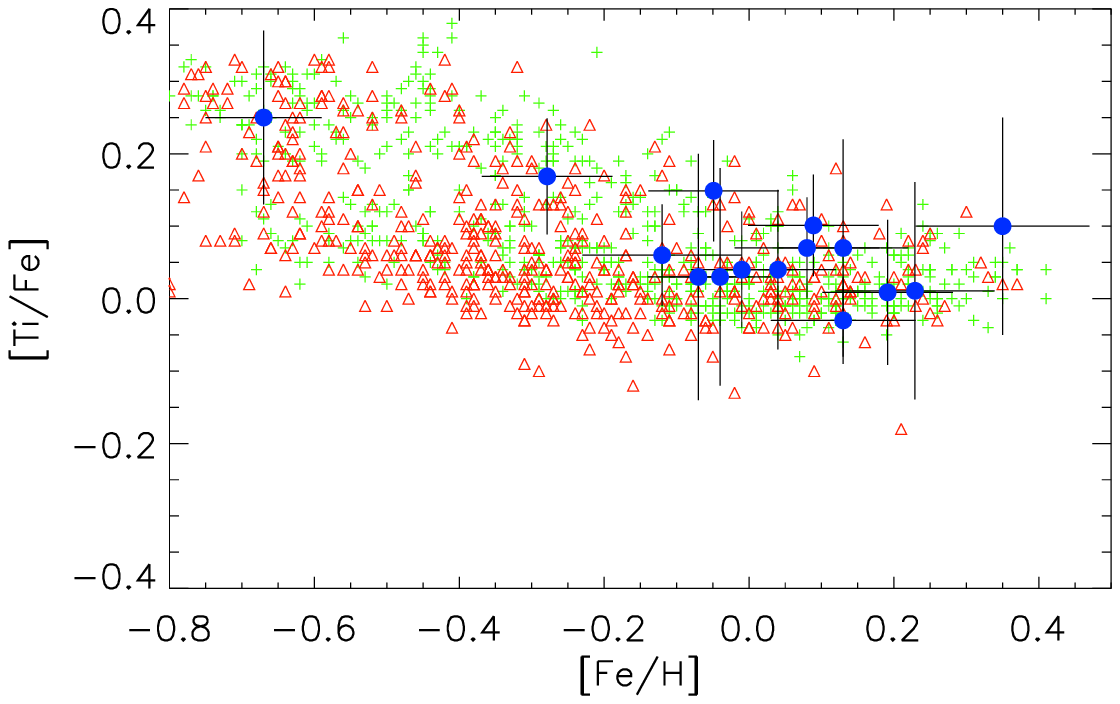}
\end{minipage}
\caption{[X/Fe] vs [Fe/H] for $\alpha$-elements(Mg, Si, Ca, Ti). Blue filled circles are member stars; Green pluses are comparison stars from \citet{ben14}; Red triangles are comparison stars from \citet{ven04}. \label{Figure 4}}
\end{center}
\end{figure}

\begin{figure}[htbp]
\begin{center}
\begin{minipage}[c]{0.45\textwidth}
\centering
\includegraphics[scale=0.6]{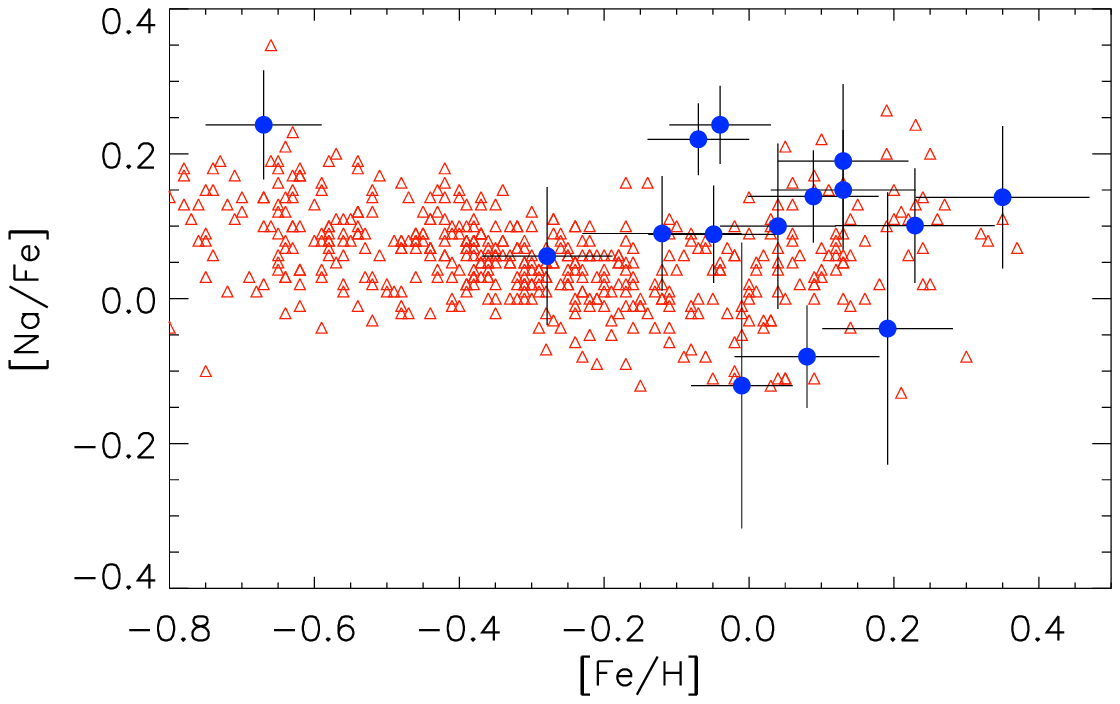}
\end{minipage}
\begin{minipage}[c]{0.45\textwidth}
\centering
\includegraphics[scale=0.6]{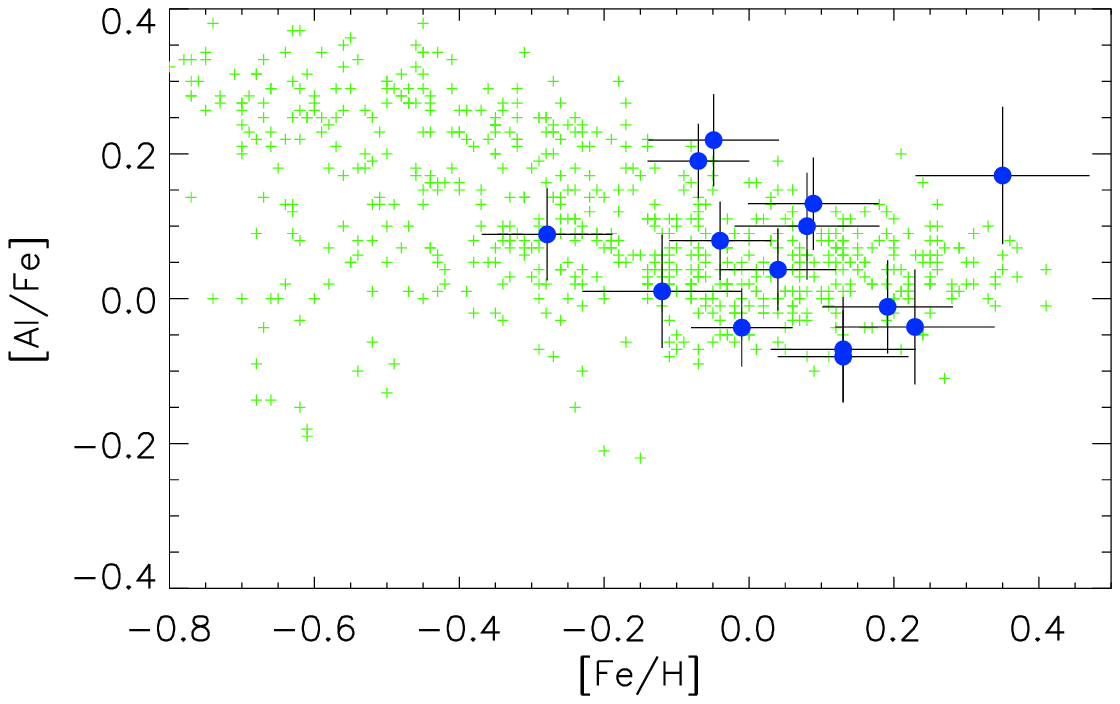}
\end{minipage}
\end{center}
\vspace{0pt}
\begin{center}
\begin{minipage}[c]{0.45\textwidth}
\centering
\includegraphics[scale=0.6]{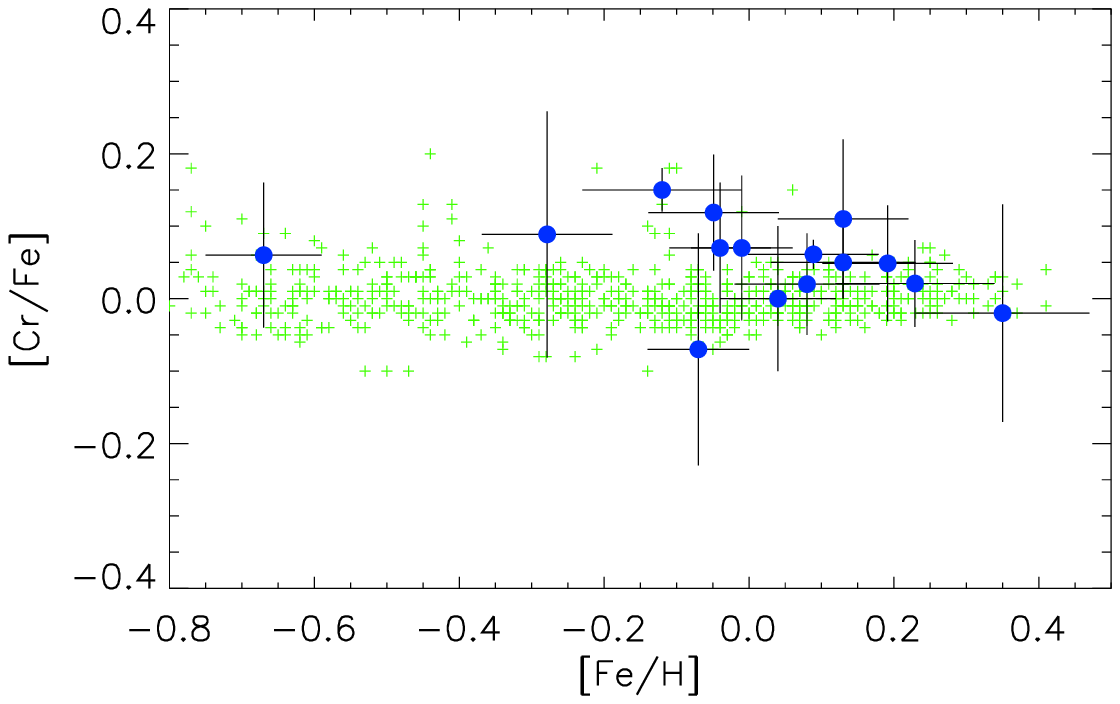}
\end{minipage}
\begin{minipage}[c]
{0.45\textwidth}
\centering
\includegraphics[scale=0.6]{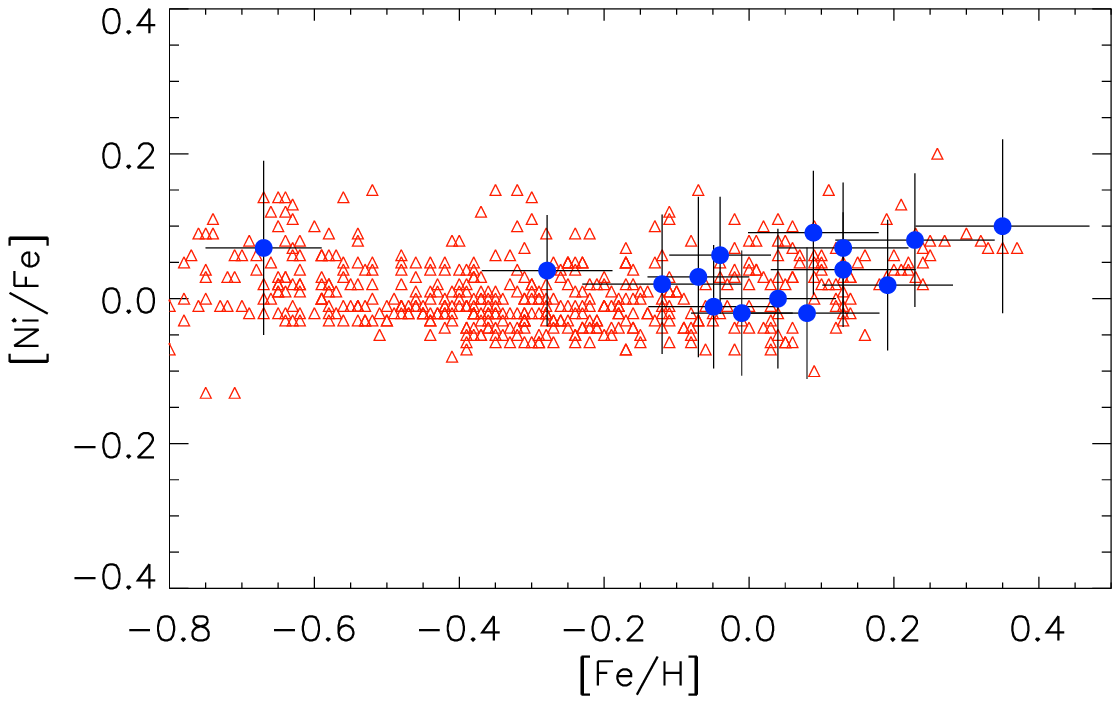}
\end{minipage}
\end{center}
\vspace{0pt}
\begin{center}
\begin{minipage}[c]{0.45\textwidth}
\centering
\includegraphics[scale=0.6]{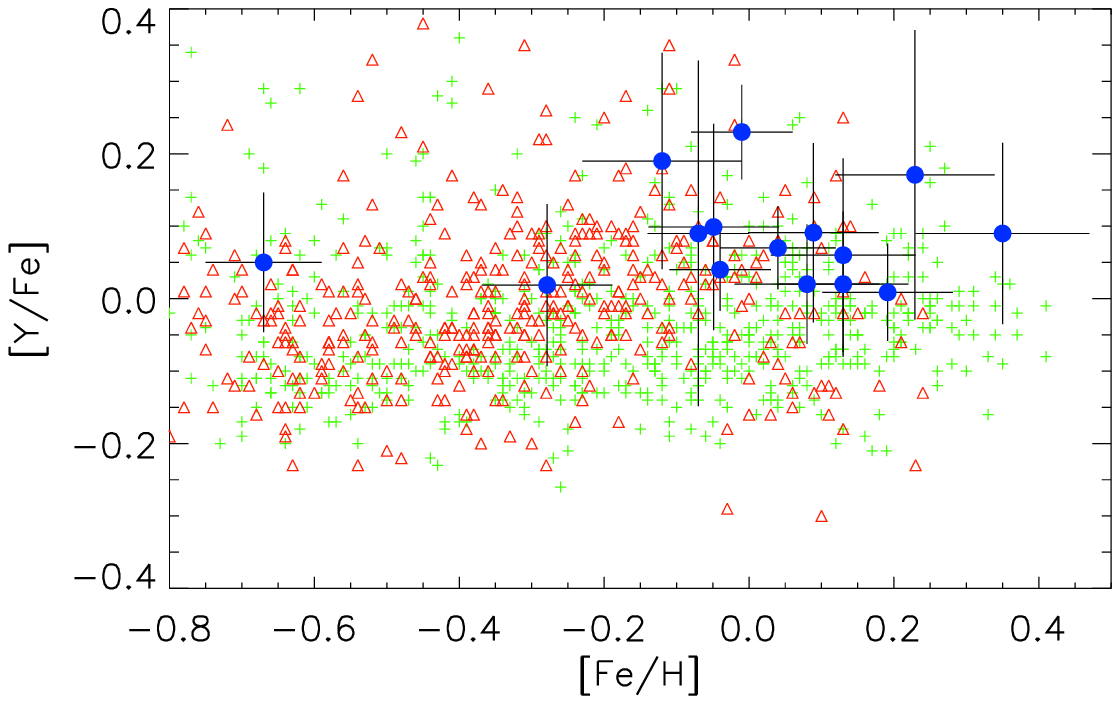}
\end{minipage}
\begin{minipage}[c]{0.45\textwidth}
\centering.
\includegraphics[scale=0.6]{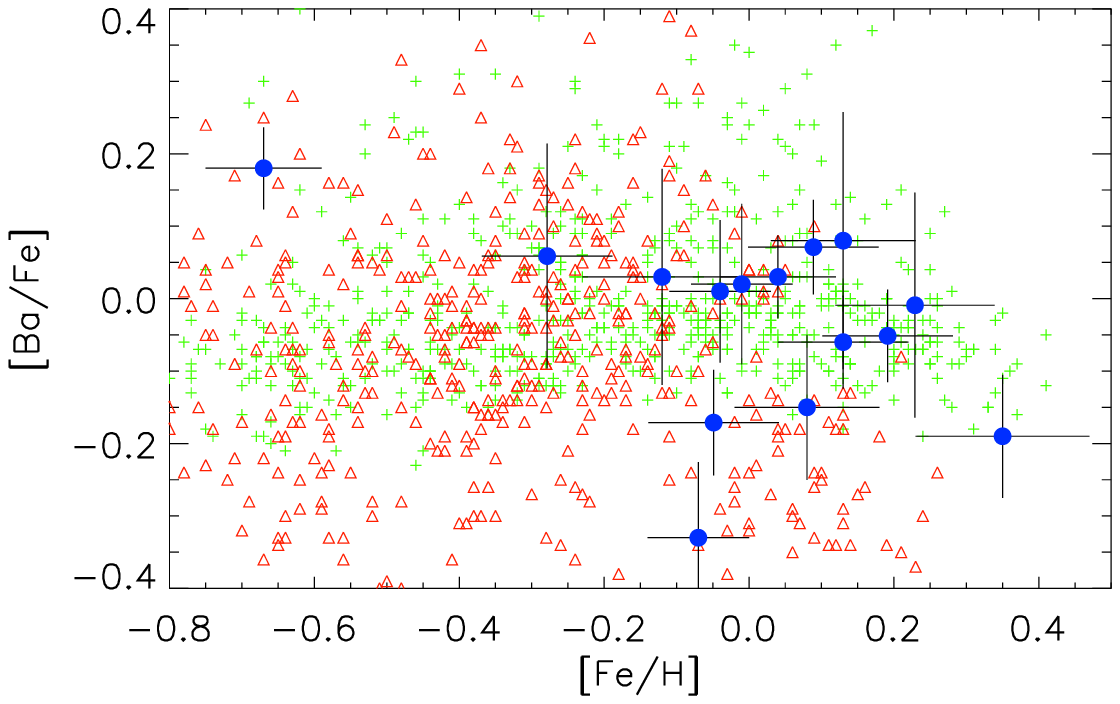}
\end{minipage}
\caption{[X/Fe] vs [Fe/H] for Fe-peak, odd-Z and s-process elements (Al, Ba, Cr, Ni, Na, Y). Blue filled circles are member stars; Green pluses are comparison stars from \citet{ben14}; Red triangles are comparison stars from \citet{ven04}. \label{Figure 5}}
\end{center}
\end{figure}

\begin{figure}
\epsscale{.70}
\plotone{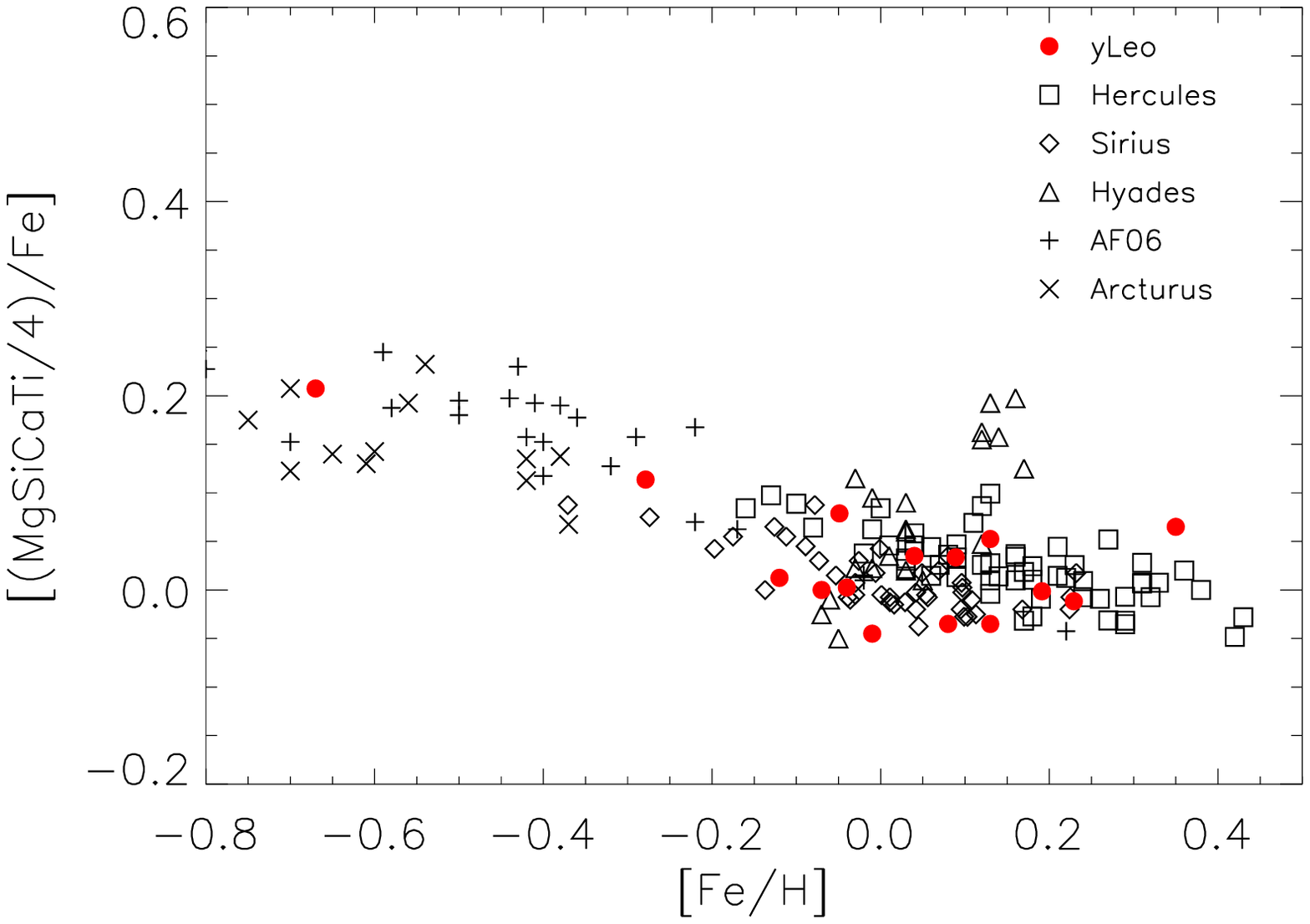}
\caption{Comparison of [$\alpha$/Fe] vs [Fe/H] between $\gamma$ Leo moving group with other moving groups. The red filled circles are member stars of  $\gamma$ Leo moving group. Black squares are member stars of Hercules moving group from \citet{ram16}. Diamond are member stars of the Sirius moving group from \citet{tab17}. Triangles are member stars of the Hyades moving group from \citet{sil11}. Pluses and crosses are member stars of the AF06 and Arcturus moving group, respectively, from \citet{ram12}. See text for details.\label{Figure 6}}
\end{figure}

$\alpha$-elements (Mg, Si, Ca, Ti) are mainly produced in Type \uppercase\expandafter{\romannumeral2} supernovae (SN \uppercase\expandafter{\romannumeral2}) nucleosynthesis, while iron is produced in both SN \uppercase\expandafter{\romannumeral2} and SN \uppercase\expandafter{\romannumeral1}a events. The [$\alpha$/Fe] ratio is a key chemical signature, because it well reflects the star formation history of the stellar system. As found in the comparision stars in solar neighbourhood, all these four $\alpha$-elements abundances of 15 member stars show decreasing trend with increasing metallicity for lower metallicities and they reach a plateau at higher metallicity \citep{edv93,che00}.

The $\alpha$-elements abundance [$\alpha$/Fe]$=$([Mg/Fe]+[Si/Fe]+[Ca/Fe]+[Ti/Fe])/4 ranges from -0.045 to 0.114. The mean and standard deviation of member stars' $\alpha$-elements abundances are 0.031 and 0.062, respectively. In figure 6, we compared [$\alpha$/Fe] with [Fe/H] of $\gamma$ Leo moving group with other known moving groups. The metallicity of the $\gamma$ Leo moving group spreads a relatively large range. The dispersion of member stars' $\alpha$-elements abundance is large too, but it's close to the Hyades moving group and AF06 moving group. At lower metallicity, the [$\alpha$/Fe] distribution of the $\gamma$ Leo moving group is similar to those of AF06 moving group, while at higher metallicity, the [$\alpha$/Fe] distribution of the $\gamma$ Leo moving group is close to the Hercules moving group.

Na and Al are odd-Z elements and thought to be produced in SN \uppercase\expandafter{\romannumeral2} and SN \uppercase\expandafter{\romannumeral1}b/c \citep{nom84}. Though it's not quit clear in Al, the Na distribution of member stars obviously shows a upturn (\citet{edv93,shi04}) in the comparision stars. Stars of the $\gamma$ Leo moving group follow this trend, although the sample size of this study is still limited. Al lines of the star J2154+1418 are so weak that they are not included in our analysis. The iron-peak elements (Cr, Ni) are believed to have the same patterns as iron. The scatters of [Cr/Fe] and [Ni/Fe] of the member stars are small as the comparision stars. Y and Ba are light and heavy neutron-capture elements, respectively. The abundances of these elements in member stars have relatively large errors, and comparision stars distribute in a wider range. According to these chemical abundances, we suggest the stars of the $\gamma$ Leo moving group are born in situ.

\section{CONCLUTION}
We have observed 18 candidates of the $\gamma$ Leo moving group members selected by the \textit{UVW} criteria from the LAMOST survey. Three stars are spectroscopic binaries and excluded from the sample. For the remaining fifteen stars, a detailed abundance analysis is carried out. The abundance pattern of member stars shows no evident difference from those of comparision stars. The large dispersion of metallicity in member stars suggests that the $\gamma$ Leo moving group is not from some chemically homogeneous origins. We suppose the $\gamma$ Leo moving group is originated from dynamical effects, perhaps related to the effect of the spiral arms.  For example, Figure 18 of \citet{ant11} shows that it is possible that spiral arms can generate a structure at this velocity region. However, small variations of the simulation parameter can produce very different velocity structures. In the future, we will do some dynamical simulations to better understand the origin the $\gamma$ Leo moving group.

The Gaia's high precision astrometric data brings great convenience to the study of moving groups in the solar neighbourhood. Chemical abundances from high resolution spectra play an important role in disentangling the degeneracy of many causes determining the local velocity structures.

\acknowledgments
We thank Bharat Kumar Yerra, Li Haining, Tan Kefeng and Liu Yujuan for their constructive suggestions and discussion. This work is supported by the Astronomical Big Data Joint Research Center, co-founded by the National Astronomical Observatories, Chinese Academy of Sciences and the Alibaba Cloud. This study is supported by the National Natural Science Foundation of China under grant No. 11390371, 11233004, U1431106, 11573035, 11625313, 11603033, the National Key Basic Research Program of China (973 program) 2014CB845701 and 2014CB845703, and JSPS - CAS Joint Research Program. WA is partially supported by JSPS KAKENHI Grant Number 16H02168.


\begin{deluxetable}{cccrrrrrrl}
\tabletypesize{\scriptsize}
\tablecolumns{10}
\tablewidth{0pc}
\tablecaption{Total sample of analysed stars. \label{tbl-1}
The columns are, respectively: identifier, right ascension, declination, velocity components \textit{U,V,W} (km s$^{-1}$), parallax (mas), proper motion components (km s$^{-1}$) in right ascension and declination, identifier names in the SIMBAD Astronomical Database.}
\tablehead{
\colhead{id} &\colhead{RA} & \colhead{Dec} & \colhead{\textit{U}}  & \colhead{\textit{V}} & \colhead{\textit{W}} & \colhead{$\pi$} & \colhead{pmra} & \colhead{pmdec} &  \colhead{simbadname}
}
\startdata
    J0158+3955&015836.1&395532&57.55&3.04 &4.12 &3.81&-26.33&-5.67 &TYC2820-1624-1\\
    J0159+2636&015939.4&263632&62.21&-1.20&2.59 &3.48&-24.79&-14.93&TYC1760-111-1\\
    J0211+4235&021144.4&423532&57.85&-2.63&6.50 &3.57&-20.40&-3.25 &TYC2838-1980-1\\
    J0216+4119&021649.4&411906&54.16&6.99 &3.29 &3.35&-21.96&-2.36 &TYC2838-96-1\\
    J0219+4506&021945.0&450638&55.98&10.35&4.43 &4.86&-34.69&1.86  &TYC3294-1261-1\\
    J0219+5623&021948.7&562344&60.69&8.76 &5.35 &6.55&-49.37&13.81 &TYC3694-1282-1\\
    J1446+2955&144637.2&295502&55.74&4.94 &4.83 &5.39&0.28  &-44.43&TYC2022-417-1\\
    J1517+1026&151719.5&102611&61.44&0.23 &-1.38&4.80&25.87 &-42.56&TYC923-1342-1\\
    J1735+2650&173524.6&265055&58.90&8.16 &2.28 &5.11&3.46  &-48.10&TYC2084-1906-1\\
    J1747+2228&174737.6&222845&57.09&-3.10&4.96 &4.28&-6.54 &-41.72&TYC1564-641-1\\
    J2101+0005&210113.2&000525&58.69&-2.84&7.28 &4.99&-38.13&-35.42&TYC526-1195-1\\
    J2154+1418&215455.0&141810&54.51&-0.34&6.71 &3.59&-27.73&-24.14&TYC1134-136-1\\
    J2208+2549&220801.3&254932&60.92&1.12 &-2.56&5.95&-46.61&-51.18&TYC2208-1077-1\\
    J2253+0315&225320.8&031554&64.54&3.18 &6.92 &4.97&-50.22&-33.95&TYC572-172-1\\
    J2328+2620&232822.8&262005&58.02&7.09 &4.72 &3.45&-31.48&-18.22&TYC2253-1606-1\\
\enddata
\end{deluxetable}

\begin{deluxetable}{cccccccccc}
\tabletypesize{\scriptsize}
\tablecolumns{10}
\tablewidth{0pc}
\tablecaption{ The basic stellar parameters for our 15 stars.\label{tbl-2}
The columns are, respectively: identifier, colour V$-$K, stellar mass, bolometric magnitude, effective temperature from photometry, effective temperature from spectroscopy, surface gravity from photometry, surface gravity from parallax, micro-turbulent velocity, signal to noise ratio at 5500 {\AA} per pixel.}
\tablehead{
\colhead{id} &  \colhead{$V-K$}&\colhead{Mass} & \colhead{M$_{bol}$} & \colhead{$T_{\rm eff,pho}$}  & \colhead{$T_{\rm eff,ab}$} & \colhead{$\log g_{pho}$} & \colhead{$\log g_{ab}$} & \colhead{$\xi$} & \colhead{S/N}\\
 & \colhead{mag} &\colhead{(M$_\bigodot$)} & \colhead{mag} & \colhead{(K)}  & \colhead{(K)} & \colhead{(km s$^{-2}$)} & \colhead{(km s$^{-2}$)} & \colhead{(km s$^{-1}$)} &

}
\startdata

    J0158+3955&1.26&1.4&3.36&6112&6389&4.14&4.23&1.4&81\\
    J0159+2636&1.51&1.1&3.86&5744&6204&4.12&4.03&1.1&72\\
    J0211+4235&1.38&1.2&3.66&5937&6351&4.14&4.25&1.2&76\\
    J0216+4119&1.48&1.1&4.38&5776&5972&4.33&4.34&0.7&60\\
    J0219+4506&1.14&1.2&4.30&6341&5848&4.52&4.57&0.4&59\\
    J0219+5623&0.48&1.2&2.68&5805&5780&4.46&4.34&0.6&123\\
    J1446+2955&1.64&0.9&5.04&5512&5597&4.43&4.42&0.5&82\\
    J1517+1026&2.10&0.8&4.68&5009&5034&4.06&4.04&0.5&50\\
    J1735+2650&1.52&1.1&4.58&5726&5616&4.40&4.30&0.5&57\\
    J1747+2228&1.68&0.9&4.67&5468&5896&4.29&4.30&0.9&80\\
    J2101+0005&1.39&1.2&3.91&5930&5645&4.25&4.24&0.7&73\\
    J2154+1418&1.06&1.1&3.94&6500&6508&4.37&4.44&1.6&69\\
    J2208+2549&1.55&0.9&4.98&5677&5688&4.49&4.47&0.5&53\\
    J2253+0315&1.40&1.1&4.47&5909&5740&4.40&4.37&0.6&40\\
    J2328+2620&1.33&1.2&3.84&6023&6197&4.25&4.22&1.1&58\\
\enddata
\end{deluxetable}

\begin{deluxetable}{cccccccccccc}
\tabletypesize{\scriptsize}
\tablecolumns{12}
\tablewidth{0pc}
\tablecaption{ The [Fe/H] and [X/Fe] abundances for our 15 stars.\label{tbl-3}}
\tablehead{
\colhead{id} &  \colhead{[Fe/H]}&\colhead{[Ca/Fe]} & \colhead{[Mg/Fe]} & \colhead{[Si/Fe]}  & \colhead{[Ti/Fe]} & \colhead{[Ba/Fe]} & \colhead{[Ni/Fe]} & \colhead{[Cr/Fe]} & \colhead{[Y/Fe]}& \colhead{[Al/Fe]}& \colhead{[Na/Fe]}
}
\startdata

J0158+3955& 0.35&  0.03&  0.05&  0.08&  0.10& -0.19&  0.10& -0.02&  0.07&  0.17&  0.14\\
J0159+2636& 0.13&  0.00&  0.09&  0.05&  0.07& -0.06&  0.07&  0.11&  0.02& -0.08&  0.19\\
J0211+4235& 0.04&  0.02&  0.04&  0.04&  0.04&  0.03&  0.00&  0.00&  0.07&  0.04&  0.10\\
J0216+4119& 0.13& -0.06& -0.05&  0.00& -0.03&  0.08&  0.04&  0.05&  0.06& -0.07&  0.15\\
J0219+4506&-0.01& -0.06& -0.13& -0.03&  0.04&  0.02& -0.02&  0.07&  0.23& -0.04& -0.12\\
J0219+5623&-0.28&  0.06&  0.09&  0.14&  0.17&  0.06&  0.04&  0.09&  0.02&  0.09&  0.06\\
J1446+2955&-0.12& -0.08& -0.01&  0.08&  0.06&  0.03&  0.02&  0.15&  0.19&  0.01&  0.09\\
J1517+1026&-0.07& -0.04& -0.16&  0.17&  0.03& -0.33&  0.03& -0.07&  0.09&  0.19&  0.22\\
J1735+2650& 0.23& -0.10&  0.00&  0.04&  0.01& -0.01&  0.08&  0.02&  0.17& -0.04&  0.10\\
J1747+2228&-0.04&  0.00& -0.08&  0.06&  0.03&  0.01&  0.06&  0.07&  0.04&  0.08&  0.24\\
J2101+0005& 0.09& -0.02& -0.04&  0.09&  0.10&  0.07&  0.09&  0.06&  0.09&  0.13&  0.14\\
J2154+1418&-0.67&  0.17&  0.22&  0.19&  0.25&  0.18&  0.07&  0.06&  0.05&   -  &  0.24\\
J2208+2549& 0.08& -0.13& -0.10&  0.02&  0.07& -0.15& -0.02&  0.02&  0.02&  0.10& -0.08\\
J2253+0315&-0.05& -0.00&  0.08&  0.09&  0.15& -0.17& -0.01&  0.12&  0.10&  0.22&  0.09\\
J2328+2620& 0.19& -0.01& -0.01&  0.01&  0.01& -0.05&  0.02&  0.05&  0.01& -0.01& -0.04\\

\enddata
\end{deluxetable}

\begin{deluxetable}{clccccccccccc}
\tabletypesize{\scriptsize}
\tablecolumns{13}
\tablewidth{0pc}
\tablecaption{The uncertainties of the abundances. \label{tbl-4}}
\tablehead{
\colhead{id} & \colhead{} & \colhead{$\Delta$[Fe]}&\colhead{$\Delta$[Ca]} & \colhead{$\Delta$[Mg]} & \colhead{$\Delta$[Si]}  & \colhead{$\Delta$[Ti]} & \colhead{$\Delta$[Ba]} & \colhead{$\Delta$[Ni]} & \colhead{$\Delta$[Cr]} & \colhead{$\Delta$[Y]}& \colhead{$\Delta$[Al]}& \colhead{$\Delta$[Na]}
}
\startdata

     & $\Delta$EW(+2.3)&  0.04&  0.04&  0.03&  0.04&  0.05&  0.05&  0.04&  0.05&  0.07&  0.04&  0.10\\
     & $\Delta$\teff (+100) &  0.07&  0.07&  0.10&  0.04&  0.09&  0.03&  0.07& -0.01&  0.04&  0.05&  0.12\\
    J0158+3955 & $\Delta\log g$(+0.12)& -0.01& -0.02& -0.02& -0.01&  0.00&  0.04& -0.01&  0.04&  0.06& -0.01&  0.06\\
     & $\Delta$[Fe/H](+0.11)& -0.00&  0.01&  0.00 & 0.01&  0.01&  0.03&  0.01&  0.03&  0.05&  0.00&  0.08\\
     & $\Delta \xi$(+0.1)& -0.03& -0.02& -0.03 &-0.01& -0.02& -0.06& -0.02& -0.02& -0.03& -0.01&  0.06\\
     & $\sigma_{total}$&  0.09&  0.09&  0.11 & 0.06&  0.11&  0.10&  0.08&  0.07&  0.12&  0.07&  0.19\\
    \hline
    & $\Delta$EW(+2.3)&  0.05&  0.04&  0.04 & 0.04&  0.06&  0.04&  0.04&  0.05&  0.05&  0.04&  0.03\\
    & $\Delta$\teff (+100) &  0.07&  0.07&  0.09 & 0.04&  0.09&  0.03&  0.07& -0.02&  0.02&  0.04&  0.05\\
    J0159+2636& $\Delta\log g$(+0.12)& -0.02& -0.02& -0.02 &-0.01&  0.00&  0.01& -0.01&  0.03&  0.04& -0.01& -0.02\\
    & $\Delta$[Fe/H](+0.11)&  0.00& -0.00&  0.01 &-0.00&  0.00&  0.03&  0.00&  0.01&  0.03& -0.01&  0.00\\
    & $\Delta \xi$(+0.1)& -0.03& -0.02& -0.02 &-0.01& -0.02& -0.06& -0.02& -0.04& -0.03& -0.01& -0.01\\
    & $\sigma_{total}$&  0.09&  0.09&  0.10 & 0.06&  0.11&  0.08&  0.08&  0.07&  0.08&  0.06&  0.06\\
    \hline
    & $\Delta$EW(+2.3)&  0.04&  0.05&  0.04 & 0.04&  0.06&  0.04&  0.05&  0.05&  0.05&  0.05&  0.04\\
    & $\Delta$\teff (+100) &  0.06 & 0.07 & 0.08  &0.04 & 0.09 & 0.05 & 0.07& -0.01&  0.02&  0.05&  0.05\\
    J0211+4235& $\Delta\log g$(+0.12)& -0.02& -0.02& -0.01 &-0.00&  0.00&  0.02& -0.01&  0.04&  0.04&  0.00& -0.02\\
    & $\Delta$[Fe/H](+0.11)& -0.01& -0.00& -0.00 &-0.00&  0.00&  0.03&  0.00&  0.01&  0.02&  0.00&  0.00\\
    & $\Delta \xi$(+0.1)&-0.03 & -0.02 &-0.02 &-0.00& -0.01& -0.05& -0.02& -0.03& -0.03&  0.00& -0.01\\
    & $\sigma_{total}$&  0.08&  0.09&  0.09 & 0.06&  0.11&  0.09&  0.09&  0.07&  0.08&  0.07&  0.07\\
    \hline
    & $\Delta$EW(+2.3)&  0.05&  0.03&  0.04 & 0.04&  0.06&  0.03&  0.04&  0.05&  0.07&  0.05&  0.03\\
    & $\Delta$\teff (+100) &  0.08&  0.07&  0.09 & 0.04&  0.10&  0.04&  0.07&  0.08&  0.02&  0.05&  0.06\\
    J0216+4119& $\Delta\log g$(+0.12)& -0.02& -0.04& -0.02 &-0.00& -0.01&  0.00& -0.02& -0.02&  0.04& -0.01& -0.03\\
    & $\Delta$[Fe/H](+0.11)&  0.01&  0.01&  0.01 & 0.01&  0.00&  0.06&  0.01&  0.01&  0.03&  0.00&  0.01\\
    & $\Delta \xi$(+0.1)& -0.03& -0.02& -0.02 &-0.00& -0.03& -0.02& -0.02& -0.01& -0.03&  0.00& -0.01\\
    & $\sigma_{total}$&  0.10&  0.09&  0.10 & 0.06&  0.12&  0.08&  0.09 & 0.10&  0.09&  0.07&  0.07\\
    \hline
    & $\Delta$EW(+2.3)&  0.05&  0.04&  0.04 & 0.04&  0.07&  0.02&  0.06 & 0.05&  0.07&  0.06&  0.04\\
    & $\Delta$\teff (+100) &  0.07&  0.07&  0.10 & 0.02&  0.11&  0.03&  0.07&  0.07&  0.01&  0.05&  0.05\\
    J0219+4506& $\Delta\log g$(+0.12)& -0.02& -0.03& -0.02 &-0.00& -0.01& -0.01&  0.00& -0.02&  0.03&  0.00& -0.03\\
    & $\Delta$[Fe/H](+0.11)&  0.01&  0.02&  0.01 & 0.01&  0.00&  0.06&  0.02&  0.01&  0.02&  0.00&  0.00\\
    & $\Delta \xi$(+0.1)& -0.01& -0.01& -0.01 &-0.01& -0.01& -0.02&  0.00& -0.01& -0.03&  0.00& -0.01\\
    & $\sigma_{total}$&  0.09&  0.09&  0.11 & 0.05&  0.13&  0.07&  0.09&  0.09&  0.08&  0.08&  0.07\\
    \hline
    & $\Delta$EW(+2.3)&  0.06&  0.04&  0.04 & 0.04&  0.08&  0.04&  0.06&  0.09&  0.11&  0.06&  0.04\\
    & $\Delta$\teff (+100) &  0.09&  0.08&  0.10&  0.03&  0.10&  0.05&  0.07&  0.07&  0.01&  0.04&  0.06\\
    J0219+5623& $\Delta\log g$(+0.12)& -0.01& -0.03& -0.02& -0.00& -0.01& -0.00& -0.01& -0.01&  0.04& -0.01& -0.02\\
    & $\Delta$[Fe/H](+0.11)&  0.01&  0.01&  0.02& -0.00&  0.00&  0.06&  0.01& -0.00&  0.03& -0.01&  0.00\\
    & $\Delta \xi$(+0.1)& -0.02& -0.01& -0.01& -0.01& -0.02& -0.02& -0.01& -0.01& -0.02&  0.00&  0.00\\
    & $\sigma_{total}$&  0.11&  0.10&  0.11&  0.05&  0.13&  0.09&  0.09&  0.11&  0.12&  0.07&  0.07\\
    \hline
    & $\Delta$EW(+2.3) & 0.05&  0.04&  0.03&  0.04&  0.06&  0.02&  0.05&  0.07&  0.09&  0.04&  0.03\\
    & $\Delta$\teff (+100)  & 0.08 & 0.08&  0.10 & 0.01 & 0.11 & 0.03&  0.06& -0.02&  0.01&  0.05&  0.06\\
    J1446+2955& $\Delta\log g$(+0.12)& -0.02& -0.03& -0.03& -0.00& -0.01& -0.01& -0.01&  0.04&  0.04& -0.01& -0.03\\
    & $\Delta$[Fe/H](+0.11)&  0.01&  0.01&  0.01&  0.01&  0.00&  0.06&  0.02&  0.03&  0.04&  0.00&  0.00\\
    & $\Delta \xi$(+0.1)&-0.02&-0.01 & -0.01& -0.01& -0.02& -0.03& -0.01& -0.01& -0.01&  0.00& -0.01\\
    & $\sigma_{total}$& 0.10&  0.10 & 0.11&  0.04&  0.13&  0.08&  0.08&  0.09 & 0.11 & 0.06&  0.07\\
    \hline

    & $\Delta$EW(+2.3)&  0.04&  0.03&  0.02&  0.04&  0.05&  0.02&  0.04&  0.04&  0.07&  0.03&  0.03\\
    & $\Delta$\teff (+100) &  0.05&  0.10&  0.10& -0.02&  0.13&  0.03  &0.03&  0.09&  0.02&  0.06  &0.08\\
    J1517+1026& $\Delta\log g$(+0.12)& -0.02& -0.05& -0.04&  0.01& -0.02& -0.01& -0.01& -0.03&  0.03& -0.02 &-0.04\\
    & $\Delta$[Fe/H](+0.11)&  0.03&  0.03&  0.03&  0.03&  0.00&  0.07&  0.03&  0.01&  0.05&  0.01&  0.02\\
    & $\Delta \xi$(+0.1)& -0.02& -0.01& -0.01& -0.01& -0.03& -0.03& -0.02& -0.02& -0.04&  0.00&  0.00\\
    J1517+1026 & $\sigma_{total}$&  0.08&  0.12&  0.11&  0.06&  0.14&  0.08&  0.06&  0.11&  0.10&  0.07&  0.10\\
    \hline

    & $\Delta$EW(+2.3)&  0.05&  0.04  &0.02&  0.04&  0.07&  0.03&  0.05&  0.05&  0.09&  0.03&  0.04\\
    & $\Delta$\teff (+100) &  0.07&  0.07  &0.06&  0.01&  0.10&  0.03&  0.06&  0.07&  0.01&  0.05&  0.06\\
    J1735+2650& $\Delta\log g$(+0.12)& -0.02& -0.03 &-0.05& -0.00& -0.01 & 0.00& -0.01& -0.02&  0.04& -0.02& -0.01\\
    & $\Delta$[Fe/H](+0.11)&  0.02&  0.02 & 0.02&  0.02&  0.01 & 0.07&  0.03&  0.01&  0.04&  0.00&  0.00\\
    & $\Delta \xi$(+0.1)& -0.02& -0.01 &-0.01& -0.01& -0.02 &-0.02& -0.02& -0.01& -0.03& -0.01& -0.01\\
    & $\sigma_{total}$&  0.09&  0.09&  0.08&  0.05&  0.12 & 0.08&  0.09&  0.09&  0.11&  0.06&  0.07\\
    \hline
    & $\Delta$EW(+2.3)&  0.05&  0.04&  0.03&  0.04&  0.09 & 0.03&  0.05&  0.06&  0.06&  0.05&  0.03\\
    & $\Delta$\teff (+100) &  0.08&  0.07&  0.09&  0.03&  0.09 & 0.04&  0.07&  0.08&  0.03&  0.05&  0.06\\
    J1747+2228& $\Delta\log g$(+0.12)& -0.01& -0.03& -0.02& -0.00& -0.01 & 0.00& -0.01& -0.01&  0.04& -0.00& -0.03\\
    & $\Delta$[Fe/H](+0.11)&  0.01&  0.00& -0.01& -0.00&  0.00&  0.05& -0.00&  0.01&  0.03& -0.00&  0.01\\
    & $\Delta \xi$(+0.1)& -0.02& -0.02& -0.02& -0.01& -0.02& -0.03& -0.02& -0.01& -0.05& -0.00& -0.01\\
    & $\sigma_{total}$&  0.10&  0.09&  0.10&  0.05&  0.13&  0.08&  0.09&  0.10&  0.10&  0.07&  0.07\\
    \hline
    & $\Delta$EW(+2.3)&  0.05&  0.04&  0.03&  0.04&  0.06&  0.03&  0.04&  0.05&  0.07&  0.04&  0.03\\
    & $\Delta$\teff (+100) &  0.08&  0.08&  0.09&  0.02&  0.11&  0.03&  0.06&  0.08&  0.01&  0.05&  0.06\\
    J2101+0005& $\Delta\log g$(+0.12)& -0.01& -0.02& -0.03&  0.01& -0.01& -0.00& -0.01& -0.02 & 0.04& -0.01& -0.02\\
    & $\Delta$[Fe/H](+0.11)&  0.02&  0.02&  0.02&  0.02&  0.01&  0.07&  0.02&  0.01 & 0.04&  0.00&  0.00\\
    & $\Delta \xi$(+0.1)& -0.02& -0.01& -0.02& -0.00& -0.02& -0.02& -0.02& -0.01 &-0.04&  0.00& -0.01\\
    & $\sigma_{total}$&  0.10&  0.09&  0.10&  0.05&  0.13&  0.08&  0.08&  0.10 & 0.10&  0.06&  0.07\\
    \hline
    & $\Delta$EW(+2.3)&  0.06&  0.03&  0.04&  0.08&  0.06&  0.04&  0.07&  0.05 & 0.07& -&  0.05\\
    & $\Delta$\teff (+100) &  0.07&  0.05&  0.07&  0.03&  0.03&  0.06&  0.06&  0.00 & 0.03& -&  0.04\\
    J2154+1418& $\Delta\log g$(+0.12)& -0.00& -0.02& -0.01& -0.00&  0.04&  0.01&  0.00&  0.04 & 0.04& -& -0.01\\
    & $\Delta$[Fe/H](+0.11)& -0.00&  0.01&  0.00& -0.00&  0.01&  0.01&  0.01&  0.00 & 0.01& -&  0.00\\
    & $\Delta \xi$(+0.1)& -0.02& -0.02& -0.01& -0.01& -0.01& -0.04&  0.00& -0.01 &-0.01& -& -0.01\\
    & $\sigma_{total}$&  0.09&  0.07&  0.08&  0.09&  0.08&  0.08&  0.09&  0.06 & 0.09& -&  0.07\\
    \hline
    & $\Delta$EW(+2.3)&  0.04&  0.03&  0.03&  0.04&  0.06&  0.04&  0.04&  0.06 & 0.07&  0.04&  0.03\\
    & $\Delta$\teff (+100) &  0.07&  0.07&  0.09&  0.02&  0.11&  0.03&  0.05&  0.08&  0.02&  0.05&  0.06\\
    J2208+2549& $\Delta\log g$(+0.12)& -0.02& -0.04& -0.03& -0.00& -0.01& -0.00& -0.01& -0.01 & 0.04& -0.01& -0.03\\
    & $\Delta$[Fe/H](+0.11)&  0.02&  0.01&  0.02&  0.02&  0.01 & 0.07&  0.02&  0.02 & 0.04&  0.00&  0.01\\
    & $\Delta \xi$(+0.1)& -0.02& -0.02& -0.02& -0.01& -0.01& -0.02& -0.02& -0.01 &-0.03& -0.01& -0.01\\
    & $\sigma_{total}$&  0.09&  0.09&  0.10&  0.05&  0.13&  0.09&  0.07&  0.10 & 0.10&  0.07&  0.07\\
    \hline
    & $\Delta$EW(+2.3)&  0.04&  0.03&  0.03&  0.04&  0.06&  0.03&  0.05&  0.06&  0.06&  0.04&  0.03\\
    & $\Delta$\teff (+100) &  0.07&  0.07&  0.10&  0.02&  0.11&  0.03&  0.07& -0.03&  0.02&  0.05&  0.07\\
    J2253+0315& $\Delta\log g$(+0.12)& -0.02& -0.04& -0.03& -0.00& -0.01& -0.00& -0.01&  0.04&  0.03& -0.01& -0.02\\
    & $\Delta$[Fe/H](+0.11)&  0.01&  0.01&  0.00&  0.01&  0.00&  0.06&  0.02&  0.02&  0.04&  0.00&  0.01\\
    & $\Delta \xi$(+0.1)& -0.02& -0.02& -0.02& -0.01& -0.02& -0.03& -0.01& -0.02& -0.04&  0.00&  0.00\\
    & $\sigma_{total}$&  0.09&  0.09&  0.11&  0.05&  0.13&  0.08&  0.09&  0.08&  0.09&  0.06&  0.08\\
    \hline
    & $\Delta$EW(+2.3)&  0.05&  0.04&  0.04&  0.04&  0.06&  0.04&  0.06&  0.05&  0.06&  0.05&  0.04\\
    & $\Delta$\teff (+100) &  0.07&  0.06&  0.09&  0.03&  0.09&  0.04&  0.07&  0.07&  0.02&  0.04&  0.05\\
    J2328+2620& $\Delta\log g$(+0.12)& -0.01& -0.03& -0.02& -0.01&  0.00&  0.01&  0.00& -0.01&  0.04& -0.01& -0.01\\
    & $\Delta$[Fe/H](+0.11)&  0.00& -0.00&  0.00& -0.00&  0.00&  0.04&  0.01&  0.00&  0.03& -0.01&  0.01\\
    & $\Delta \xi$(+0.1)& -0.03& -0.02& -0.02& -0.01& -0.01& -0.05& -0.01& -0.02& -0.03& -0.01& -0.01\\
    & $\sigma_{total}$&  0.09&  0.08&  0.10&  0.05&  0.11&  0.09&  0.09&  0.09&  0.09&  0.07&  0.07\\

\enddata
\end{deluxetable}


\begin{thebibliography}{}

\bibitem[Alonso et al (1995)]{alo95}Alonso A., Arribas S., Martinez-Roger C., 1995, A\&A, 297, 197
\bibitem[Alonso et al (1996)]{alo96}Alonso A., Arribas S., Martinez-Roger C., 1996, A\&A, 313, 873
\bibitem[Alonso et al (1999)]{alo99}Alonso A., Arribas S., Martinez-Roger C., 1999, A\&AS, 140, 261
\bibitem[Antoja et al (2011)]{ant11}Antoja T., Figueras F., Romero-G\'omez M., et al., 2011, MNRAS, 418(3): 1423
\bibitem[Antoja et al (2012)]{ant12}Antoja T., Helmi A., Bienayme O., et al., 2012, \mnras, 426,L1.
\bibitem[Aoki et al (2005)]{aok05}Aoki W., Honda S., Beers. T., et al., 2005, ApJ, 632, 611
\bibitem[Asplund et al (2009)]{asp09}Asplund M., Grevesse N., Sauval A. J., et al., 2009, \araa, 47: 481
\bibitem[Bensby (2007)]{ben07}Bensby T., Zenn A. R., Oey M. S., \& Feltzing S., 2007, \apj, 663,L13
\bibitem[Bensby et al (2014)]{ben14}Bensby T., Feltzing S., Oey M. S., 2014, A\&A, 562, A71
\bibitem[Chen et al (2000)]{che00}Chen Y. Q., Nissen P. E., Zhao G., Zhang H. W., Benoni T., 2000, A\&AS, 141, 49
\bibitem[Cui et al (2012)]{cui12}Cui X., Zhao Y., Chu Y., et al., 2012, RAA, 12, 1197
\bibitem[De Silva et al (2007)]{sil07}De Silva G. M., Freeman K. C., Bland-Hawthorn J., et al., 2007, AJ, 133, 694
\bibitem[De Silva et al (2011)]{sil11}De Silva G. M., Freeman K. C., Bland-Hawthorn J., et al., 2011, MNRAS, 415(1): 563
\bibitem[De Simone et al (2004)]{sim04}De Simone, R., Wu, X., \& Tremaine, S., 2004, MNRAS, 350, 627
\bibitem[Dehnen (2000)]{deh00}Dehnen W., 2000, \aj, 119, 800
\bibitem[Edvardsson et al. (1993)]{edv93}Edvardsson B., Andersen J., Gustafsson B., Lambert D. L., Nissen P. E., Tomkin J., 1993, A\&A, 275, 101
\bibitem[Eggen (1959a)]{egg59a}Eggen O. J., 1959, The Observatory, 79, 88
\bibitem[Eggen (1959b)]{egg59b}Eggen O. J., 1959, The Observatory, 79, 182
\bibitem[Eggen (1958-1998)]{egg58}Eggen O. J., 1958, MNRAS, 118, 154\\
    Eggen O. J., 1965, Observatory, 85, 191\\
    Eggen O. J., 1969, PASP, 81, 553\\
    Eggen O. J., 1970, PASP, 82, 99\\
    Eggen O. J., 1971, PASP, 83, 271\\
    Eggen O. J., 1974, PASP, 86, 162\\
    Eggen O. J., 1977, ApJ, 215, 812\\
    Eggen O. J., 1978a, ApJ, 222, 191\\
    Eggen O. J., 1978b, ApJ, 222, 203\\
    Eggen O. J., 1983, AJ, 88, 813\\
    Eggen O. J., 1992a, AJ, 104, 1906\\
    Eggen O. J., 1992b, AJ, 104, 1482\\
    Eggen O. J., 1996, AJ, 111, 1615\\
    Eggen O. J., 1998, AJ, 115, 2453\\
\bibitem[Famaey et al (2005)]{fam05}Famaey B., Jorissen A., Luri X., et al., 2005, A\&A, 430, 165
\bibitem[Freeman \& Bland-Hawthorn (2002)]{fre02}Freeman, K., \& Bland-Hawthorn, J. 2002, ARA\&A, 40, 487
\bibitem[Fux (2001)]{fux01}Fux, R. 2001, \apj, 373, 511
\bibitem[Gaia Collaboration et al (2016a)]{bro16}Brown A. G. A., Vallenari A., Prusti T., et al., 2016, A\&A, 595: A2
\bibitem[Gaia Collaboration et al (2016b)]{pru16}Prusti T., De Bruijne J. H. J., Brown A. G. A., et al., 2016, A\&A, 595: A1
\bibitem[Green et al (2015)]{gre15}Green G. M., Schlafly E. F., Finkbeiner D. P., et al., 2015, \apj, 810(1): 25
\bibitem[Huang et al (2015)]{hua15}Huang Y., Liu X. W., Yuan H. B., et al., 2015, MNRAS, 449(1): 162
\bibitem[Kalnajs (1991)]{kal91}Kalnajs, A. 1991, in Dynamics of Disc Galaxies, ed. B. Sundelius (G¨oteborg: Department of Astronomy and Astrophysics, G¨oteborg Univ.), 323
\bibitem[Klement et al (2008)]{kle08}R. Klement, B. Fuchs, and H.-W. Rix, 2008, \apj, 685, 261.
\bibitem[Kong et al (2017)]{kon17}Kong, X. M., Kumar, Y. B., Zhao, G., et al. 2017, MNRAS, 474, 2129
\bibitem[Kurucz (1993)]{kur93}Kurucz R., 1993, ATLAS9 Stellar Atmosphere Programs and 2 km/s grid. Kurucz CD-ROM No. 13. Cambridge, Mass.: Smithsonian Astrophysical Observatory, 1993., 13
\bibitem[Liang et al (2017)]{lia17}Liang X. l., Zhao .J, Oswalt T. D., et al., 2017, \apj, 844(2), 152
\bibitem[Monari et al (2017)]{mon17}Monari, G., Kawata, D., Hunt, J. A., \& Famaey, B., 2017, MNRAS, 466, L113
\bibitem[Noguchi et al (2002)]{nog02}Noguchi K., Aoki W., Kawanomoto S., et al., 2002,  PASJ, 54(6): 855.
\bibitem[Nomoto et al (1984)]{nom84}Nomoto K., Thielemann F.-K., Yokoi K., 1984, ApJ, 286, 644
\bibitem[Nordstr\"om et al(2004)]{nor04}Nordstr\"om B., Mayor M., Andersen J., et al. 2004, A\&A, 418, 989
\bibitem[P\'{e}rez-Villegas et al(2017)]{per17}P\'{e}rez-Villegas, A., Portail, M., Wegg, C., \& Gerhard, O., 2017, ApJL, 840, L2
\bibitem[Quillen \& Minchev (2005)]{qui05}Quillen A. C., \& Minchev I., 2005, AJ, 130, 576
\bibitem[Ramya et al (2012)]{ram12}Ramya P., Reddy B. E., Lambert D. L., 2012, MNRAS, 425(4): 3188
\bibitem[Ramya et al (2016)]{ram16}Ramya P., Reddy B. E., Lambert D. L., et al., 2016, MNRAS, 460(2): 1356
\bibitem[Schlafly \& Finkbeiner (2011)]{sch11}Schlafly \& Finkbeiner 2011, ApJ, 737, 103
\bibitem[Shi et al (2011)]{shi04}Shi J. R., Gehren T., Zhao G., 2004, A\&A, 423(2): 683
\bibitem[Shkolnik et al (2012)]{shk12}Shkolnik E. L., Anglada-Escud\'e G., Liu M. C., et al., 2012, \apj, 758(1): 56
\bibitem[Skuljan et al (1997)]{sku97}Skuljan, J., Cottrell, P. L., \& Hearnshaw, J. B. 1997, in Hipparcos—Venice ’97, ed. B. Battrick (ESA SP-402; Noordwijk: ESA), 525
\bibitem[Tabernero et al (2017)]{tab17}Tabernero H. M., Montes D., Hern\'andez J. G., et al., 2017, A\&A, 597: A33
\bibitem[Venn et al (2004)]{ven04}Venn K. A., Irwin M., Shetrone M. D., et al., 2004, AJ, 128(3), 1177
\bibitem[Wenger et al (2000)]{sbd00}Wenger, M., Ochsenbein, F., Egret, D., et al. 2000, A\&AS, 143, 9
\bibitem[Wielen (1971)]{wie71}Wielen R., 1971, A\&A, 13, 309
\bibitem[Yi et al (2001)]{yi01}Yi S. K., Demarque P., Kim Y. C., et al., 2001, ApJS, 136(2): 417
\bibitem[Yi et al (2003)]{yi03}Yi S. K., Kim Y. C., Demarque P., 2003, ApJS, 144, 259
\bibitem[Zhao et al (2006)]{zhao06}Zhao G., Chen Y. Q., Shi J. R., et al., 2006, ChJAA , 6(3): 265
\bibitem[Zhao et al (2007)]{zhao07}Zhao J. K., Zhao G., Chen, Y. Q., 2007, PNAOC, 4, 153
\bibitem[Zhao et al (2009)]{zhao09}Zhao J.K, Zhao G, \& Chen Y.Q 2009, ApJ, 692, L113
\bibitem[Zhao et al (2012)]{zhao12}Zhao G., Zhao Y. H., Chu Y. Q., et al. 2012, RAA, 12,723

\end{thebibliography}
\end{document}